\documentclass[10pt,journal,compsoc]{IEEEtran}
\usepackage{cite}
\usepackage{amsmath,amssymb,amsfonts}
\usepackage{algorithmic}
\usepackage{graphicx}
\usepackage{textcomp}
\usepackage{xcolor}
\usepackage{color}
\usepackage{multirow}
\usepackage{caption}
\usepackage{subcaption}
\usepackage{listings}
\usepackage{todonotes}
\usepackage{booktabs}
\usepackage[ruled,vlined,linesnumbered]{algorithm2e}
\def\BibTeX{{\rm B\kern-.05em{\sc i\kern-.025em b}\kern-.08em
    T\kern-.1667em\lower.7ex\hbox{E}\kern-.125emX}}

\newcommand{\tool}{Finch}
\newtheorem{example}{Example}

\begin{document}

\title{\tool: Fuzzing with Quantitative and Adaptive \\ Hot-Bytes Identification }
\author{Tai D. Nguyen, Long H. Pham, Jun Sun}

\IEEEtitleabstractindextext{
\begin{abstract}
Fuzzing has emerged as a powerful technique for finding security bugs in complicated real-world applications. American fuzzy lop (AFL), a leading fuzzing tool, has demonstrated its powerful bug finding ability through a vast number of reported CVEs. However, its random mutation strategy is unable to generate test inputs that satisfy complicated branching conditions (e.g., magic-byte comparisons, checksum tests, and nested if-statements), which are commonly used in image decoders/encoders, XML parsers, and checksum tools. Existing approaches (such as Steelix and Neuzz) on addressing this problem assume unrealistic assumptions such as we can satisfy the branch condition byte-to-byte or we can identify and focus on the important bytes in the input (called hot-bytes) once and for all. In this work, we propose an approach called \tool~which is designed based on the following principles. First, there is a complicated relation between inputs and branching conditions and thus we need not only an expressive model to capture such relationship but also an informative measure so that we can learn such relationship effectively. Second, different branching conditions demand different hot-bytes and we must adjust our fuzzing strategy adaptively depending on which branches are the current bottleneck. We implement our approach as an open source project and compare its efficiency with other state-of-the-art fuzzers. Our evaluation results on 10 real-world programs and LAVA-M dataset show that \tool~achieves sustained increases in branch coverage and discovers more bugs than other fuzzers. 
\end{abstract}}
\maketitle
\IEEEdisplaynontitleabstractindextext
\IEEEpeerreviewmaketitle
\section{Introduction}

Fuzzing has emerged as a powerful technique for finding security bugs in complicated real-world applications. It is used to find many reported CVEs in all sort of projects from compilers (e.g., gcc and clang), network tools (e.g., tcpdump and openssh), GNU softwares (e.g., readelf and objdump) to image processing libraries (e.g., libtiff and libpng). While the vast majority of the discovered bugs cause denial of service to commonly used software, some severe bugs such as remote code execution and privilege escalation are also detected.

In general, there are many types of fuzzing, e.g., whitebox fuzzing, blackbox fuzzing, and greybox fuzzing. Among them, coverage-guided greybox fuzzing (CGF) is the most successful one. Its goal is to find test inputs that maximize code coverage within a given amount of time. Among many developed CGF tools, american fuzzy lop (AFL) is a leading fuzzer. It employs genetic algorithms to mutate user-provided test inputs and execute them to discover buggy behaviours. However, CGF fuzzers heavily rely on random mutations. This often prevents them from bypassing complicated branching conditions and thus leaves many vulnerabilities unreachable.

\definecolor{solarized@base03}{HTML}{002B36}
\definecolor{solarized@base02}{HTML}{073642}
\definecolor{solarized@base01}{HTML}{586e75}
\definecolor{solarized@base00}{HTML}{657b83}
\definecolor{solarized@base0}{HTML}{839496}
\definecolor{solarized@base1}{HTML}{93a1a1}
\definecolor{solarized@base2}{HTML}{EEE8D5}
\definecolor{solarized@base3}{HTML}{FDF6E3}
\definecolor{solarized@yellow}{HTML}{B58900}
\definecolor{solarized@orange}{HTML}{CB4B16}
\definecolor{solarized@red}{HTML}{DC322F}
\definecolor{solarized@magenta}{HTML}{D33682}
\definecolor{solarized@violet}{HTML}{6C71C4}
\definecolor{solarized@blue}{HTML}{268BD2}
\definecolor{solarized@cyan}{HTML}{2AA198}
\definecolor{solarized@green}{HTML}{859900}

\lstset{language=C++,
        basicstyle=\footnotesize\ttfamily,
        numbers=left,
        numberstyle=\footnotesize,
        tabsize=2,
        breaklines=true,
        escapeinside={@}{@},
        numberstyle=\tiny\color{solarized@base01},
        keywordstyle=\color{solarized@green},
        stringstyle=\color{solarized@cyan}\ttfamily,
        identifierstyle=\color{solarized@blue},
        commentstyle=\color{solarized@base01},
        emphstyle=\color{solarized@red},
        frame=single,
        rulecolor=\color{black},
        rulesepcolor=\color{black},
        showstringspaces=false,
        linewidth=0.45\textwidth
}
\begin{figure}[h!]
    \centering
\begin{lstlisting}
u8 input[8];
// checksum test using nested if-statements
if (sum(input)<8)
  if(input[sum(input)]==0x01)
    bug(1);
// magic-byte comparison
if (u32(input)*2==0xdeadbeee))
    bug(2);
\end{lstlisting}
    \caption{Magic-byte comparisons and checksum tests}
    \label{fig:complicated_conditions}
\end{figure}

\bigbreak
\noindent\emph{Common problems in fuzzing}. Complicated branching conditions (e.g., magic-byte comparisons, checksum tests, and nested if-statements) are common obstacles that prevent CGF fuzzers from exploring more branches. An example of such conditions is shown in Figure \ref{fig:complicated_conditions} where bug \#1 is protected by a checksum and bug \#2 is covered if the first 4 bytes of the provided input equal to a magic number $0x6f56df77$. It is obvious that covering these branches is non-trivial for CGF fuzzers as the probability of generating an exact combination of 4 bytes is negligible. To overcome such problems, different techniques have been proposed \cite{li2017steelix, rawat2017vuzzer, she2020mtfuzz, gan2020greyone, chen2018angora, stephens2016driller, chen2020savior}. For instance, taint tracking and symbolic execution models have been applied to track the flow of input to the branching conditions. By doing that, hot-bytes are identified and then mutated more often than other bytes. 

To detect hot-bytes but avoid unacceptable overhead of dynamic taint tracking or symbolic execution, existing approaches often propose their own lightweight taint algorithms. Notably, Angora \cite{chen2018angora} and GREYONE \cite{gan2020greyone} record byte offsets which flow into each unsolved branching condition. REEDQUEEN \cite{aschermann2019redqueen} identifies input segments that are directly used in branching conditions. However, the lightweight taint algorithms are not able to capture the relation between inputs and branching conditions quantitatively (e.g., only byte offsets are returned). As a result, no clue on how to mutate hot-bytes is given. With the capability of capturing complicated dependency, deep neural networks have been recently considered as a complementary choice to tainting analysis. Neuzz \cite{she2019neuzz} detects hot-bytes using a trained neural network. However, its approach is not adaptive as the guidance (i.e., gradient) quickly becomes outdated after several branches are covered.

\bigbreak
\noindent\emph{Our approach}. In this paper, we propose a different fuzzing scheme. Let us assume that we are trying to bypass the true branch of $u32(input)*2$==$0xdeadbeee$ given an initial test input $t$. A traditional fuzzer will repeatedly mutate $t$ until one of its permutation $t'$ satisfies the branch. Furthermore, even if hot bytes (the first four bytes) are identified, existing approaches have to exhaustively try every possible value in the worst case to cover the branch. In contrast, we use an expressive model (a neural network) to identify the hot-bytes and furthermore, we measure how far a test $t$ is from covering the branch with a distance. In detail, we also mutate $t$, but $t$ is replaced with $t'$ if $|u32(t')*2 - 0xdeadbeee| < |u32(t)*2 - 0xdeadbeee|$. The comparison means that $t'$ is better than $t$. With this approach, we gradually narrow down the difference between $u32(input)*2$ and $0xdeadbeee$, and eventually bypass the targeted branch. 

In general, we aim to systematically answer the following questions and design the fuzzer accordingly:
\begin{itemize}
    \item \emph{Q1: How to quantitatively compare $t$ and $t'$ in term of which one is more likely to cover a branch?} Inspired by search-based software testing \cite{panichella2017automated, mcminn2004search, harman2009theoretical, nguyen2020sfuzz}, we propose a branch distance $f_n(t)$, which measure how far a branch $br_n$ from being covered by a test input $t$. If $f_n(t') < f_n(t)$ then $t'$ is better than $t$. In the above example, $f_n(t) = |u32(t)*2 - 0xdeadbeee|$
    \item \emph{Q2: How to adaptively minimize fuzzing queue to reduce the number of redundant mutations?} Existing approaches either use simple heuristics like keeping those tests which just cover some new branches as seeds (without considering whether covering those branches is meaningful for covering other uncovered branches), or keep mutating the same hot-bytes~\cite{she2019neuzz}. Such approaches are unlikely effective as during fuzzing, the branches to be covered keeps changing and thus the relevant seeds and the mutation strategy must be adaptive as well. To overcome the weakness above, we propose the min-Pareto set which is a lightweight fuzzing queue. If $br_n$ is one of the targeted branches and $f_n(t')$ is minimal then $t'$ belongs to the min-Pareto set.
\end{itemize}

We combine the answers of Q1 and Q2 in~\tool, a smart adaptive fuzzer. We compare \tool~with 6 state-of-the-art fuzzers including AFL~ \cite{afl}, FairFuzz~\cite{lemieux2018fairfuzz}, Neuzz~\cite{she2019neuzz}, MOPT~\cite{lyu2019mopt}, Angora~\cite{chen2018angora}, and TortoiseFuzz~\cite{wang2020not}. We conduct multiple experiments on LAVA-M and a set of test subjects including the latest versions of 10 high-profile real-world applications covering 7 different file formats, i.e., {\it jpg}, {\it png}, {\it xml}, {\it bz2}, {\it png}, {\it tiff}, and {\it ttf}. The experiment results show that \tool~on average covers more branches and discovers many more unique crashes than all other fuzzers.

In summary, we make the following technical contributions:
\begin{itemize}
    \item Our first contribution is an algorithm for fuzzing queue minimization which eliminates redundant mutations.
    \item Our second contribution is a novel machine learning approach for identifying hot-bytes, quantitatively and adaptively.
    \item Our third contribution is an open-source fuzzer, named \tool, which is systematically evaluated on extensive experiments.
\end{itemize}

The remainder of the paper is organized as follows. Section \ref{sect:pre} provides relevant background on fuzzing, with a review of key innovations of AFL. Section \ref{sect:our_approach} presents our fuzzing approach. Section \ref{sect:implementation} shows implementation details. Section \ref{sect:evaluation} reports evaluation results. Section \ref{sect:related_work} reviews related work and Section \ref{sect:conclusion} concludes the paper.
\section{Preliminary}
\label{sect:pre}
In this section, we define the fuzzing problem and review existing most relevant grey-box fuzzing approaches.

\subsection{Problem definition}
Let $P$ be the program under test; and $B$ be the set of basic blocks (i.e., non-branching block of codes) in $P$.
Formally, a program can be represented as a \emph{control flow graph}, in which each node represents a basic block and each edge (a.k.a branch) represents a transition between 2 blocks.
Ideally, as a testing engine, a fuzzer's goal is to generate a test set which can expose all bugs in the program.
However, since we do not know where the bugs are, normally a fuzzer aims to generate a test set with high code coverage.
This is based on the observation that {\it increasing code coverage often results in more bugs found}~\cite{lemieux2018fairfuzz}.
Among different types of code coverage (e.g., block coverage and path coverage), edge coverage is the most popular choice, as it balances between effectiveness and efficiency.

In this work, we define a branch $e$ in the program as a pair of blocks $(x, y)$ with $x, y \in B$. 
Let $E$ is the set which contains all possible branches of the program $P$.
A test input $t$ is said to cover the branch $e = (x, y)$ iff its execution has a transition from $x$ to $y$.
We define a set of branches that are covered by a test input $t$ as $E_t$.
Our problem is then defined formally as follows.
\textit{Given a program P and a set of seed tests $T_0$,
generate a new set of test inputs $T$ which maximizes the number of covered branches $|C|$ in the program, where $C = \cup_{t \in T}(E_t)$, and do so as early as possible.}\\

\subsection{AFL overview}

Among multiple developed fuzzers, AFL is an industrial standard one that employs genetic algorithms to maximum code coverage. Many fuzzers are built on top of or inspired by AFL. In the following, we briefly introduce AFL by explaining its key components.  
\bigbreak
\noindent\emph{Fuzzing procedure}. The main workflow of AFL is shown in Algorithm \ref{algo:afl_loop}.
The inputs of the procedure are the program under test $P$, the set of original user-provided test inputs $T_0$, and the expected fuzzing time.
In general, AFL generates new test inputs by mutating a set of existing ones, called \textit{seed pool}.
Initially, the seed pool contains only the user-provided test inputs (line 1).
During each iteration, AFL picks a random test input from the seed pool (line 5), mutates it to generate new test inputs (line 6), and runs the mutated inputs with the target program (line 7).
The execution is monitored by the fuzzer to collect necessary information.
If the test input execution causes some buggy behaviours (e.g., makes the program crash or hang), the test input is added into a \textit{crash pool} and is used to debug the program later (lines 8-9).
If the test input executes some new branches in the program, it is added into the seed pool and will be mutated to generate more test inputs in the next iteration (lines 10-11).
In case the test input does not cause any buggy behaviours nor cover any new branch, it is deemed irrelevant and simply discarded.
The process continues until the time budget is exhausted.
At that time, all the test inputs in the seed pool and the crash pool are returned. The output of the procedure is a set of test inputs, each of which covers some unique branches in the program or causes some buggy behaviours. 

\begin{algorithm}[t]
let $seed\_pool \gets T_0$\;
let $crash\_pool \gets \emptyset$\;
\While{in time limit}{
    let $new\_seeds \gets \emptyset$\;
    \ForEach{$t$ in $seed\_pool$} {
        \ForEach{$t'$ in $mutate(t)$}{
            let $r \gets execute(t')$\;
            \If{$r$ causes buggy behaviours} {
                $crash\_pool.add(t')$\;
            }
            \ElseIf{$r$ obtains new coverage} {
                $new\_seeds.add(t')$\;
            }
        }
    }
    $seed\_pool \gets seed\_pool \cup new\_seeds$\;
}
\Return{$seed\_pool$, $crash\_pool$};
\caption{$fuzz(P, T_0, time)$}
\label{algo:afl_loop}
\end{algorithm}

\bigbreak
\noindent\emph{Mutation stages}. A test input is a sequence of bytes. In AFL, each test input is mutated in two stages: {\it deterministic} stage and {\it havoc} stage.
Mutations are based on a set of different byte-level mutation operators. Currently, there are 11 mutation operators \cite{lyu2019mopt} in AFL. In the deterministic stage, the mutation operators are applied deterministically based on a predefined algorithm. For instance, given a test input, the {\it bit\_flip} operator is applied before the {\it byte\_flip} operator. Therefore, the mutation results are identical given the same input in the deterministic stage.
Due to this reason, deterministic mutations only need to be applied once for the same test input.
So a test input is passed through the deterministic stage iff it has not been mutated previously.
In contrast, the havoc stage randomly applies mutation operators on the test inputs. Thus given the same input, it can generate different new mutants and can be invoked multiple times. Additionally, the havoc stage often changes the size of a test input by adding or deleting multiple block of bytes.

As mutation strategies have a huge impact on AFL's effectiveness, extensive studies have been conducted on improving AFL's mutation strategy. It can be improved in multiple ways, e.g., using dictionary collected from static analysis~\cite{li2017steelix}, using symbolic execution to solve constraints~\cite{chen2018angora}, and analyzing data/control dependency at run-time~\cite{gan2020greyone}. However, the more intelligent the improvement is, the more run-time information needed and thus the more overhead introduced into the fuzzer. Therefore, how to design a mutation strategy which is both effective and efficient is still an ongoing research problem.

\bigbreak
\noindent\emph{Edge coverage}. From Algorithm~\ref{algo:afl_loop}, it can be noted that discovering test inputs with new coverage is essential for AFL's performance. When no new branches are covered, AFL becomes more or less a simple random test input generation engine.
To collect the coverage information, AFL injects a small piece of code that captures the control flow of the target program. The injected code associate each basic block with a random numeral identifier.
The identifier of the branch $(x, y)$ is then computed from the identifiers of blocks $x$ and $y$ as follows:
\begin{equation*}
    \mathrm{ID}_{(x, y)} = (\mathrm{ID}_x \gg 1) \oplus \mathrm{ID}_y
    \label{eq:hash}
\end{equation*}
where $\gg$ and $\oplus$ are right-shift and xor operators respectively.
AFL keeps track of coverage branches in a bitmap and uses the above formula to compute the key value for a branch access.
Initially, the values of all the keys are unset.
Then in the execution of each test input, AFL monitors the transitions between the basic blocks in the program.
Whenever there is a transition from a block $x$ to another block $y$, the according branch identifier $\mathrm{ID}_{(x, y)}$ is computed.
In case the value at the key $\mathrm{ID}_{(x, y)}$ is unset, it is now set and the test input is deemed to cover a new branch $(x,y)$.\\
\section{Our approach}
\label{sect:our_approach}
To bypass complicated branching conditions, we build a fuzzing tool named \tool. The novelty of \tool~is twofold. First, \tool~identifies hot-bytes with an expressive neural network model. Second, \tool~measures the effectiveness of mutations on the hot-bytes quantitatively by maintaining a lightweight seed pool, called \emph{min-Pareto set}. The min-Pareto set is adaptive as it contains the seeds which have one-to-one correspondence to the branches to be covered next. This strategy helps us focus more on the test inputs which have potential to uncover new branches. Moreover, in our neural network model, test inputs chosen from the min-Pareto set are labeled with the distances measures, instead of coarse-grained coverage information. We will give the details about branch distance, how we choose the test inputs to mutate, how we train the model and use it to predict hot-bytes in subsequent sections.

\subsection{Branch distance}
\begin{figure}[h!]
    \centering
    \includegraphics[trim={0cm 7cm 10cm 0cm},clip,page=8,width=0.45\textwidth]{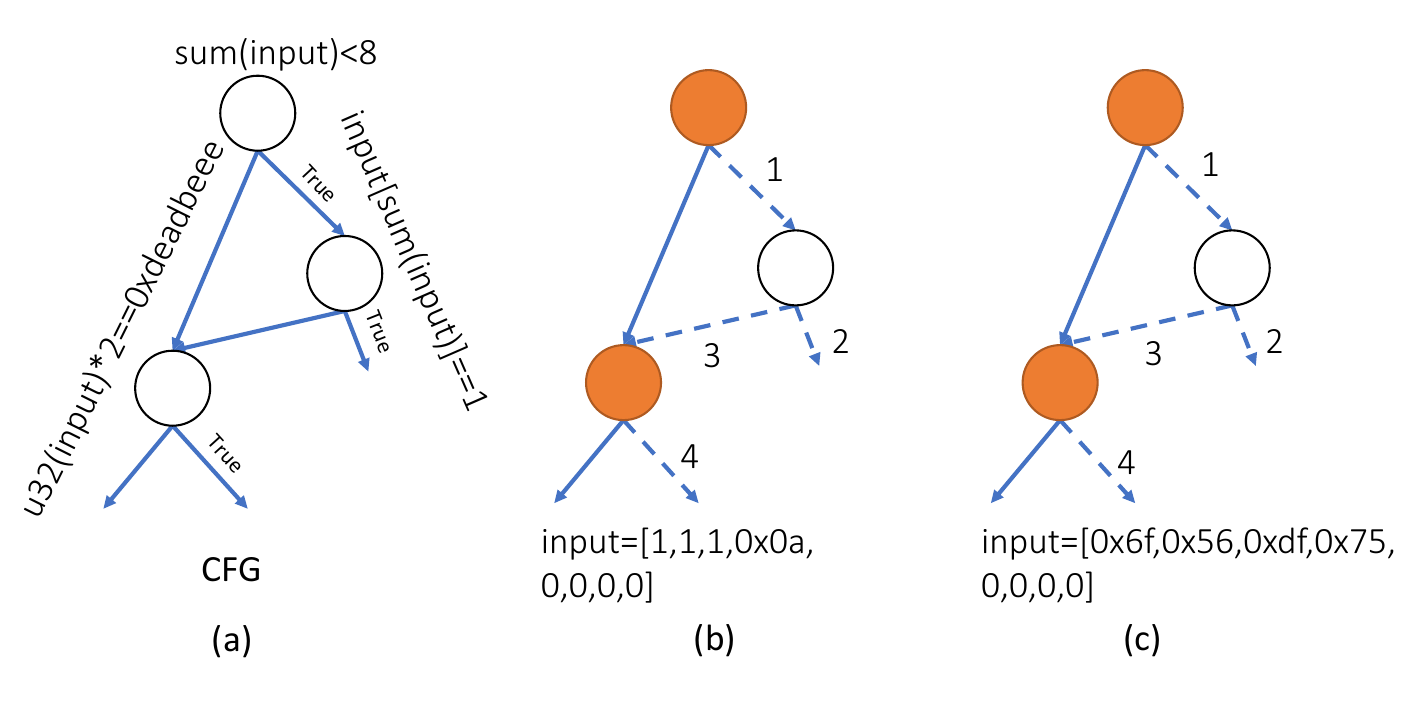}
    \caption{An example to demonstrate how a branch distance bitmap is computed}
    \label{fig:example_branch_bitmap}
\end{figure}
Inspired by search-based software testing \cite{panichella2017automated, mcminn2004search, harman2009theoretical, nguyen2020sfuzz}, we compute the distance between each test input and uncovered branches in the program. The distance often contains two components: approach distance and branch distance. Intuitively, the approach distance and branch distance measure the gap between two execution paths and two branches of a node respectively. As the concrete control flow graph of a target program is explored on the fly, it is expensive to maintain approach distance \cite{nguyen2020sfuzz} and thus, we focus on the branch distance (so as to avoid overwhelming computational overhead). In the following, we define our customized branch distance.

Let $f_n(t)$ be a function to measure how far the test input $t$ is from covering an uncovered branch $br_n$. Let us assume that $br_n$ is an outgoing uncovered branch from a branching node $n$ and $n$ is labeled with a condition $c$. At the byte-code level, $c$ is one of the following forms $true$, $false$, $a > b$, $a \ge b$, $a \le b$, $a < b$, $a = b$, $a \neq b$ where $a$ and $b$ are either a constant or a variable. The function $f_n(t)$ is then computed as follows
\begin{equation*}
    f_n(t)=\left\{
                \begin{array}{ll}
                |a - b| & \mbox{if $t$ visits node $n$ }\\
                K & \mbox{otherwise}
                \end{array}
              \right.
\end{equation*}

\noindent where $K$ is a constant representing the maximum distance.
If $c$ is either $true$ or $false$ then $|a - b|$ is 0. The function $f_n(t)$ is defined such that the closer the branch is from being covered, the smaller the resultant value is. 


For a program with $n$ uncovered branches, we define the branch distance bitmap of a test input $t$ as follows.
\[
f(t) = \{f_1(t), f_2(t), \dots, f_n(t)\}
\]

\begin{example}
Fig.~\ref{fig:example_branch_bitmap}(a) shows the control flow graph of the program shown in Fig.~\ref{fig:complicated_conditions}. Suppose we execute the program with 2 different test inputs, $t_0$ with $input$=[1,1,1,0x0a,0,0,0,0] and $t_1$ with $input$=[0x6f, 0x56,0xdf,0x75,0,0,0,0]. The execution paths of $t_0$ and $t_1$ are shown in Fig.~\ref{fig:example_branch_bitmap}(b) and~\ref{fig:example_branch_bitmap}(c), where visited nodes are highlighted in orange; dashed and solid lines indicate uncovered and covered branches respectively. With these 2 test inputs, there are 4 uncovered branches left in the program, which are marked with the number $1$, $2$, $3$, and $4$. Applying the above formula, we have $f_1(t_0)$ = $|13 - 8|$ = $5$ and $f_4(t_0)$ = $|0x0101010a * 2 - 0xdeadbeee|$ = $3702242522$. The value of $f_2(t_0)$ and $f_3(t_0)$ are $K$ because $t_0$ does not visit the node with branch $2$ and $3$ as its outgoing branches.

So the branch distance bitmap of $t_0$ is:
\begin{align*}
    f(t_0) = \{f_1(t_0), f_2(t_0), f_3(t_0), f_4(t_0)\} \\
    = \{5, K, K, 3702242522\}
\end{align*}

Similarly, the branch distance bitmap of $t_1$ is:
$$f(t_1) = \{f_1(t_1), f_2(t_1), f_3(t_1), f_4(t_1)\} = \{529, K, K, 4\}$$
\end{example}

\subsection{Choosing test inputs}

We interpret the problem of fuzzing a program as a multi-objectives optimization problem, which can be formulated as
\begin{equation*}
    \min_{t \in T} (f_1(t), f_2(t), \dots, f_n(t))
    \label{eq:moo}
\end{equation*}
    
\noindent where $f_n(t)$ returns the quantitative distance between a test input $t$ and an uncovered branch $br_n$; $T$ is a set of all possible test inputs.
In general, there does not typically exist a test input $t$ that minimizes all distances simultaneously.
So instead, we accept a test input $t$ as a solution as long as it is Pareto optimal, which is determined by a dominance.

In detail, a feasible solution $t_1$ is said to dominate another solution $t_2$ iff
\begin{itemize}
    \item[(1)] For all objectives, $t_1$ is not worse than $t_2$. Formally, $\forall i \in \{1, 2, \dots, n\} \cdot f_i(t_1) \le f_i(t_2)$.
    \item[(2)] There exists an objective that $t_1$ is strictly better than $t_2$. Formally, $\exists j \in \{1, 2, \dots, n\} \cdot f_j(t_1) < f_j(t_2)$.
\end{itemize}

With the above definition, a solution $t$ is a Pareto optimal solution if there does not exist another solution that dominates it. A set of Pareto optimal solutions lie on a Pareto boundary. 

A naive strategy is to simply select the test inputs which satisfy the Pareto optimal conditions to mutate in each generation. However, a known problem \cite{panichella2017automated} for such a strategy is that the number of Pareto optimal test inputs soars after a few generations. To overcome the problem, we apply 2 rules to reduce the number of test inputs in the Pareto boundary. The remaining test inputs after filtering are called min-Pareto set and are used to mutate.

The rules are designed as follows:

\begin{itemize}
    \item[(r1)] Removing `future' objectives: let $br_k$ be an outgoing uncovered branch of a node $k$ in a control flow graph, an objective $f_k$ is removed from our list of objectives if the node $k$ is never visited. Intuitively, because node $k$ is never visited, covering $br_k$ is not an immediate objective yet (rather covering some branches leading to node $k$ is). Furthermore, since node $k$ is never visited, we are not able to compute the distance measure and the model is unlikely able to predict meaningfully either. We call the branches according to the remaining objectives after filtering {\it just-missed branches}.
    
    \item[(r2)] Removing `redundant' test inputs: a Pareto optimal test input $t$ is removed from a set of Pareto optimal test inputs $O$ if $\forall i \in \{1, 2, \dots, n\}; \exists t' \in O; t' \neq t; f_i(t') \le f_i(t)$. 
    Intuitively, we remove a test input $t$ if for all objectives, there always exists some other test inputs which are better than $t$. We notice that the Pareto boundary is not changed this way.
\end{itemize}

Ideally, we want to find a set with a minimum number of test inputs which satisfy the above 2 rules. 
However, due to the complexity of rule r2, instead of finding the optimal set of test inputs, we use a greedy algorithm that approximates a min-Pareto set $S$ from the original Pareto set $O$ in order to reduce computational resources.
The overall workflow is shown in Algorithm \ref{algo:min_pareto_set}. Let $J_t$ be the set of objectives $f_n$ such that $\forall t' \in T$; $t' \neq t$; $f_n(t) \le f_n(t')$. The main idea of our algorithm is to sort test inputs $t \in O$ in a descending order of $|J_t|$ (line 2) and then check the test input one by one according to the sorted list (lines 3-4). For each $t$, if $J_t \neq \emptyset$ then $t$ is added to $S$ (lines 5-6). After that, for all remaining test inputs $t'$ in the list, we update $J_{t'} = J_{t'} \setminus J_{t}$ (lines 7-8) before checking the next element in the next iteration. Since $O$ is sorted only once, our algorithm only returns an approximated result for the min-Pareto set with complexity $|O|^2$.\\

\begin{algorithm}[t]
\small
let $S \gets \emptyset$\;
let $sorted\_list \gets descending\_sort(O)$\;
\While{$sorted\_list \neq \emptyset$} {
    $(t, J_t) \gets sorted\_list.pop\_front()$\;
    \If{$J_t \neq \emptyset$} {
        $S.add(t)$ \;
        \ForEach{$(t', J_{t'})$ in $sorted\_list$}{
            $J_{t'} \gets J_{t'} \setminus J_t$\;
        }
    }
}
\Return $S$\;
\caption{$min\_pareto\_set(O)$}
\label{algo:min_pareto_set}
\end{algorithm}

\begin{figure}[h!]
\centering
        \includegraphics[trim={0cm 9cm 21cm 0cm},clip,page=9,width=0.35\textwidth]{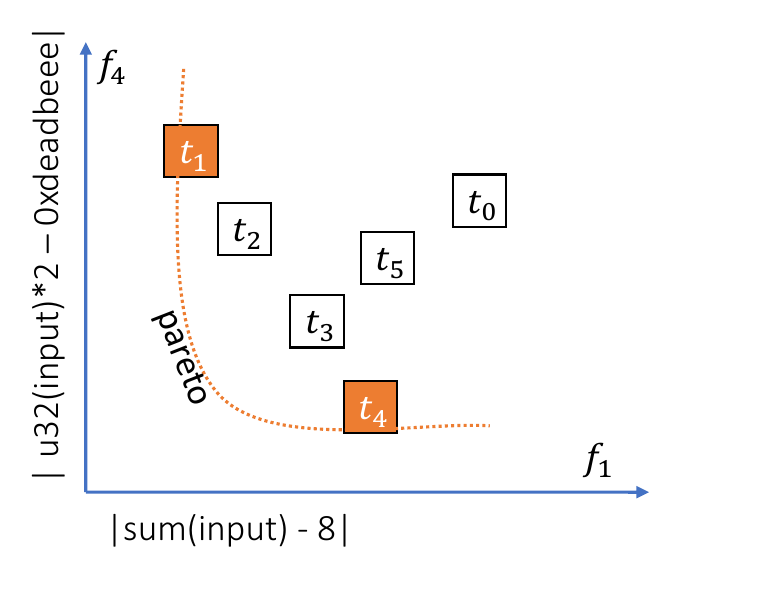}
        \caption{Two objectives optimization for the code in Fig.~\ref{fig:complicated_conditions}}
        \label{fig:vulerable_code_optimization}
\end{figure}

\begin{example}
Fig.~\ref{fig:vulerable_code_optimization} illustrates the simplified multiple-objectives problem from Fig.~\ref{fig:complicated_conditions}, in which we only consider 2 conditions at line 3 and 7. The x-axis and the y-axis represent objective functions $f_{1}$ and $f_{4}$ respectively. According to the returned values of $f_{1}$ and $f_{4}$, a test input $t$ occupies a position $(f_{1}(t), f_{4}(t))$ on $xy$-plane. Suppose $t_0$ is the original test inputs and the mutation creates test inputs from $t_1$ to $t_5$. In case the new test inputs do not cover new branches, other fuzzers (e.g., AFL and Angora) simply discards all of them  and starts a new mutation process with $t_0$ again. Our approach, on the other hand, computes the distances between the new test inputs and the uncovered branches. We then detect that $t_1$ and $t_4$, although not cover new branches, are on the Pareto boundary and are better solutions than $t_0$. So instead of discarding all generated test inputs, we update the seed pool by adding Pareto optimal test inputs $t_1$, $t_4$. After running Algorithm~\ref{algo:min_pareto_set}, while $t_0$ is removed from the seed pool, $t_1$ and $t_4$ are kept and ready to be mutated to generate new test inputs. So although no new coverage is obtained, the seed pool makes some steps toward uncovered branches. The strategy is repeated until both branches are covered.
\end{example}

\subsection{Hot-bytes prediction}

For each generation, the test inputs in the min-Pareto set are used for mutation. However, unlike AFL, where all bytes of the test inputs are mutated equally, we only focus on mutating hot-bytes of each test input. Intuitively, hot-bytes should be selected such that mutating them is more likely to cover new branches (than mutating other bytes). 

Our idea is to predict the hot-bytes using a neural network model. The novelty of our approach is the training data and the labels. First, instead of using all generated test inputs, we only use the test inputs in the min-Pareto set, which is chosen based on Algorithm~\ref{algo:min_pareto_set}. This approach helps to reduce the overhead in the training process, while still keeping most informative and updated training data. Second, instead of using the coarse-grained coverage information as the label, our model is trained with the branch distance bitmaps of the test inputs. With the training data, our model can predict which bytes need to be mutated to decrease the values of the distances between test inputs and uncovered branches. 

The number of neurons at the input layer is equal to the number of bytes of the longest test input in the training data. The number of neurons at the output layer is equal to the number of just-missed branches in the program. The test inputs are normalized by padding zero to guarantee all the test inputs have the same length as the longest one. The branch distance bitmap is also normalized by dividing by $K$ so that the branch distance bitmap has values ranging from 0 to 1. 

In the training process, we use binary cross-entropy to compute the loss between a predicted branch distance bitmap $\hat{y}$ and the ground truth $y$ = $f(t)/ K$. It is defined as follows.
\[
\mathrm{loss}(y, \hat{y}) = -\frac{1}{n}\sum_{i = 0}^n(y_i \times \log(\hat{y}_i) + (1 - y_i) \times \log(1 - \hat{y_i}))
\]
Note that $y_i = 1$ represents the maximum branch distance, which is incomparable with other values by nature. Therefore, we simply set $\hat{y_i}$ = $y_i$ if $y_i$ = $1$ to discard the loss between $y_i$ and $\hat{y_i}$.

After having the model, we can use it to predict hot-bytes of the test inputs. In particular, each test input in the min-Pareto set is fed into the model. After training, we can compute the gradients of each byte in the test input according to the model prediction. 

\begin{example}
Suppose that we want to mutate a test input $t$ with $input$=[1,1,1,0x0a,0,0,0,0]. Our trained model receives $t$ and outputs gradient scores $g$=[0.5,0.5,0.5,0.9,0,0,0,0]. The magnitude of $g$ means that the first 4 bytes are more important than others. The sign of $g$ means that we should decrease the values of the first 4 bytes to cover new branches.
\end{example}

\subsection{Hot-bytes mutation}

\begin{algorithm}[h!]
let $T \gets \emptyset$\;
let $g \gets model(t)$\;
\ForEach{range in [(0, 2), (2, 4), (4, 8), $\dots$]}{
    let $locs \gets top(g, range)$\;
    \ForEach{$dir$ in $[-1,1]$}{
        $t' \gets t.copy()$\;
        \While{$\exists x \in t'[locs]; x \neq 0 \land x \neq 255 $}{
            $t'[locs] \gets t'[locs] + dir * signs(g[locs])$\;
            $T.add(t')$\;
        }
    }
}
\Return{$T$};

\caption{$mutate\_hot\_bytes(t, model)$}
\label{algo:neu_mutate}
\end{algorithm}

Given a test input $t$, in this section, we will explain how $t$'s hot-bytes are mutated. The mutation algorithm is firstly introduced by \cite{she2019neuzz}. It is summarized in Algorithm \ref{algo:neu_mutate}. Since it is impossible to exhaustively mutate all bytes in $t$, the key idea is that we consider a group of bytes as a single byte and exhaustively mutate them. The more important bytes is grouped into a smaller group and less important bytes are grouped into a larger group (line 3-4). Each byte in a group is increased or decreased according to the sign of its gradient (line 8).

First, we compute the gradient scores $g$ using the trained model (line 2). Then, each group of sorted bytes are repeatedly selected for mutation until all bytes are grouped (line 3-4). For each group of selected bytes, we gradually increase or decrease their values by 1 (line 8) until all of them reach their boundaries (line 7). Note that function $sign(x)$ returns 1 if $x > 0$, 0 if $x = 0$ and -1 if $x<0$ and thus, $signs(g[locs])$ returns a list of $sign(x)$ where $x \in g[locs]$.

\begin{example}
Suppose that we want to mutate a test input $t$ with $input$=[1,1,1,0x0a,0,0,0,0]. Our trained model receives $t$ and outputs gradient scores $g$=[0.5,0.5,0.5,0.9,0,0,0,0]. According to algorithm \ref{algo:neu_mutate}, $t$ is mutated as follows:
\begin{itemize}
    \item Going up (dir=1,range=(2,4)): input is added to the corresponding signs. The possible permutations of input are: [1,1,\textbf{2,0x0b},0, 0,0,0], [1,1,\textbf{3,0x0c} ,0,0,0,0],$\cdots$,[1,1,\textbf{0xf6,0xff} ,0,0,0,0], $\cdots$, [1,1, \textbf{0xff,0xff} ,0,0,0,0]
    \item Going down (dir=-1,range(2,4)): input is subtracted to the corresponding signs. The possible permutations of input are: [1,1, \textbf{0,0x09},0,0,0,0], [1,1,\textbf{0,0x08} ,0,0,0,0],$\cdots$,[1,1,\textbf{0,0x06},0,0,0,0], $\cdots$, [1,1,\textbf{0,0},0,0,0,0]
\end{itemize}
\end{example}

\subsection{System overview}

\begin{algorithm}[h!]
\small
\colorbox{orange} {let $seed\_pool \gets min\_pareto\_set(T_0)$}\;
let $crash\_pool \gets \emptyset$\;
\While{in time limit}{
    \colorbox{orange} {
    let $model \gets train\_model(seed\_pool)$}\;
    let $new\_seeds \gets \emptyset$\;
    \ForEach{$t$ in $seed\_pool$}{
        \colorbox{orange} { let $T \gets mutate\_hot\_bytes(t, model) \cup havoc(t)$}\;
        \ForEach{$t'$ in $T$}{
            let $r \gets execute(t')$\;
            \If{$r$ causes buggy behaviours} {
                $crash\_pool.add(t')$\;
            }
            \ElseIf{$r$ is on Pareto boundary} {
                $new\_seeds.add(t')$\;
            }
        }
    }
    $seed\_pool \gets seed\_pool \cup new\_seeds$\;
    \colorbox{orange} {$seed\_pool \gets min\_pareto\_set(seed\_pool)$}\;
}
\Return{$seed\_pool$, $crash\_pool$};
\caption{$hot\_fuzz(P, T_0, time)$}
\label{algo:test_input_generation}
\end{algorithm}

We are now ready to present the overall workflow of \tool~shown in Algorithm \ref{algo:test_input_generation}. The difference between \tool~and AFL is highlighted in orange. Initially, we compute the min-Pareto set seed pool according to the original user-provided test inputs (line 1).
While time is available during each iteration, we train a model according to the current test inputs in seed pool and their branch distance bitmaps (line 4).
After having the model, in the mutation stage, we use Algorithm~\ref{algo:neu_mutate} to predict the hot-bytes and mutate each test input in the seed pool to get the new test inputs.
Based on the execution of the new test inputs, they may be added into the crash pool (lines 10-11) or the new seeds (lines 12-13). Note that unlike AFL, the new test input does not need to cover any new branch to be added into {\it new\_seeds}.
After mutating all the test input in the current seed pool, 
all the test inputs in {\it new\_seeds} are added into {\it seed\_pool} (line 14) and then we apply the function to find the min-Pareto set in {\it seed\_pool} again (line 15) before beginning the new iteration.
Like AFL, the output of \tool~is the set of test inputs in {\it seed\_pool} and {\it crash\_pool} (line 16).


\section{Implementation}
\label{sect:implementation}
\begin{figure}[t]
    \centering
    \includegraphics[trim={0 13cm 16cm 0cm},clip,page=2,width=0.45\textwidth]{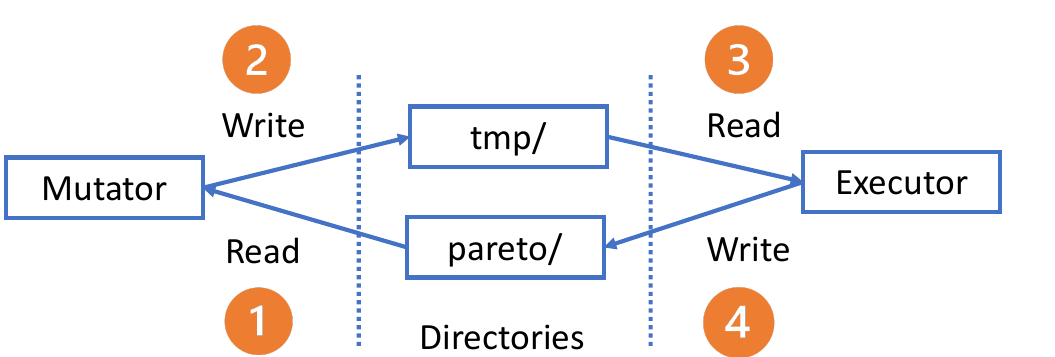}
    \caption{The modular architecture of \tool}
     \label{fig:modular_architecture}
\end{figure}

\tool~is implemented in C and Python with an estimated 4k lines of code. It has two main components: \textit{mutator} and \textit{executor}. While Python is adopted to implement a {\it mutator} due to the ease of neural network implementation using PyTorch, we choose C to implement the \textit{executor} to minimize the execution time and reuse some components of AFL. Fig.~\ref{fig:modular_architecture} presents the general workflow of \tool. The \textit{mutator} evolves the Pareto optimal test inputs located at the directory \textit{pareto/}. After that, it writes generated test inputs to the directory \textit{tmp/} and asks the \textit{mutator} to execute them. If an executed test input is Pareto optimal, the \textit{executor} sends it back to the directory \textit{pareto/}. Otherwise, it is deleted. The cycle continues until \tool~reaches predefined time limit (e.g., 24h). Note that test inputs located at directory \textit{pareto/} are Pareto optimal and they must go through the Algorithm \ref{algo:min_pareto_set} implemented in \textit{mutator} to form a min-Pareto set.
\begin{figure*}[t!]
    \centering
    \includegraphics[width=\textwidth]{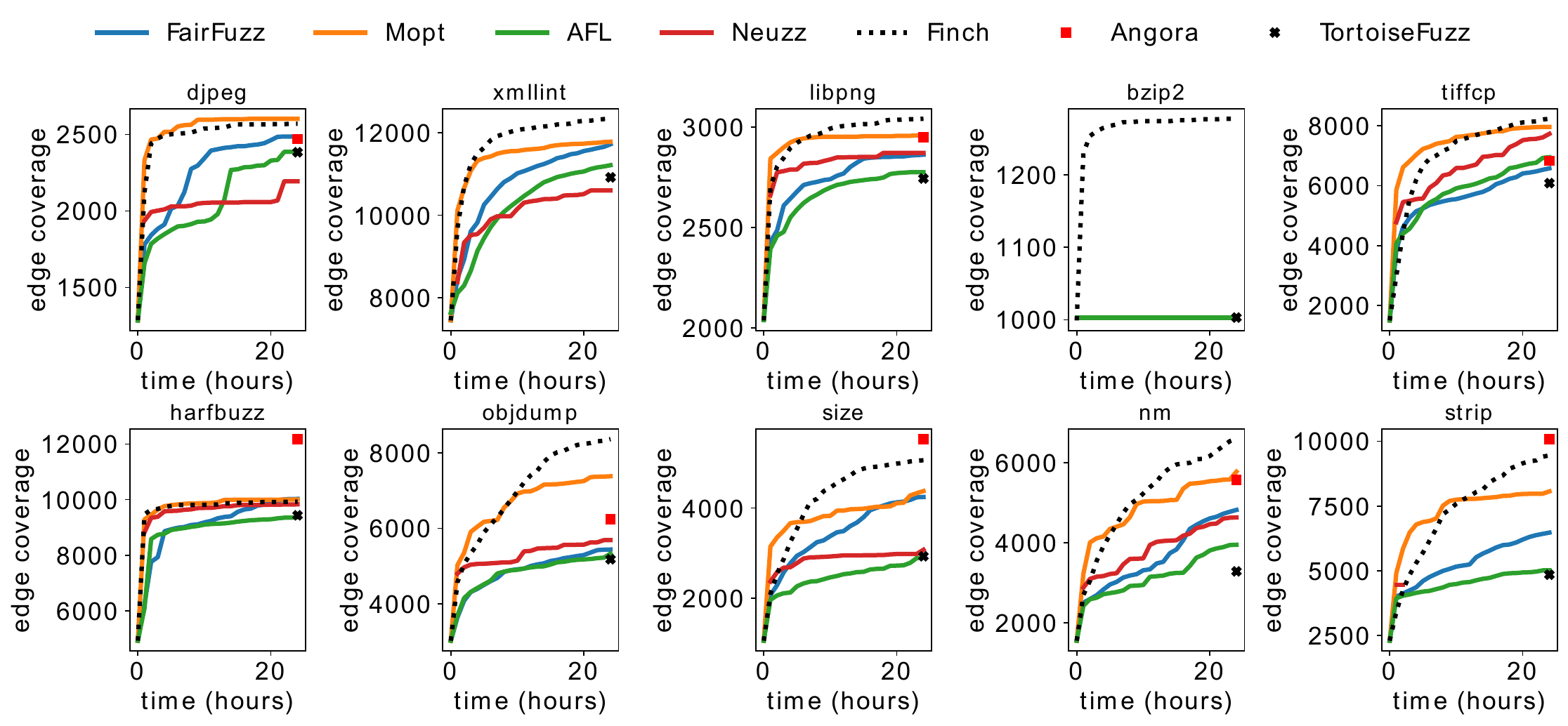}
    \caption{Edge coverage of different fuzzers running for 24 hours}
    \label{fig:edge_coverage}
\end{figure*}
\noindent\emph{Neural Network Architecture.} Our neural network is implemented in PyTorch version 1.6.0. It has one hidden fully-connected layer. The input and output layers use activation function ReLU and Sigmoid respectively. A neural network is trained for 200 epochs and deleted after assisting the gradient-guided mutation. Since our neural network is simple, the training time is often less than 1 minute. A feed-forward neural network is preferred to keep \tool~efficient and relatively explainable.

\bigbreak
\noindent\emph{Instrumentation.} To construct a branch-distance bitmap from a test input, we write an LLVM Pass to inject our handlers, which capture operands of conditional instructions, i.e., \textit{BranchInst} and \textit{SwitchInst} at run-time. Due to many different data types used in a comparison, for any comparison that we are unable to derive branch distance, the default distance 0 is used. More specifically, we maintain another shared memory along with AFL's coverage bitmap to save branch distances of both uncovered and covered branches. From these two bitmaps, we are able to infer which branch is uncovered and its corresponding distance. Note that to avoid instrument overhead, instead of making a function call to compute branch distance $|a - b|$ for every comparison instruction, we use single instruction \textit{xor} to compute distance, i.e., $a \oplus b$. The xor distance guarantees that if $a$ equals to $b$ then branch distance is $0$. Although the \textit{xor} operator is used, our metric provides greater granularity (i.e., the returned value ranges from 0 to $2^n - 1$ where $n$ is the number of bits representing data) compared to CMP feedback \cite{padhye2019fuzzfactory,LibFuzzer} (i.e., the returned value ranges from 0 to $n$)
\section{Evaluation}
\label{sect:evaluation}


In this section, we evaluate \tool~through multiple experiments on two different types of datasets: 10 real-world applications and LAVA-M \cite{li2017steelix} benchmark with injected bugs. The experiments are designed to answer the following research questions (RQ).
\begin{itemize}
    \item \textit{RQ1: Can \tool~achieve higher edge coverage than existing fuzzers?}
    \item \textit{RQ2: Is the branch-distance bitmap useful?}
    \item \textit{RQ3: How min-Pareto set help increase edge coverage?}
    \item \textit{RQ4: Can \tool~find more bugs than existing fuzzers?}
    \item \textit{RQ5: What is the overhead of training model?}
\end{itemize}

We compare \tool~with 6 state-of-the-art fuzzers on the 10 real-world applications, as shown in Table \ref{tab:experiment_setup}. While MOPT and FairFuzz achieve relatively high edge coverage in many evaluations \cite{yue2020ecofuzz,fuzzbench}, Angora and Neuzz focus on mutating hot-bytes, which is similar to \tool. TortoiseFuzz~\cite{wang2020not} has been published recently. The initial test inputs are collected from their repositories with the file size limit of 10 kb. All experimental results reported below are obtained on an Ubuntu Ubuntu 16.04 LTS machine with Intel Core(TM) i9 and 32GB of memory.

\begin{table}[t]
    \parbox{.5\textwidth}{
        \begin{tabular}{ll}
        \toprule
        \textbf{Package} & \textbf{Program} \\
        \midrule
        \multirow{4}{*}{binutils-2.36}
        & \textit{objdump -D @@} \\ 
        & \textit{nm-new -C @@} \\ 
        & \textit{size @@} \\
        & \textit{strip-new @@ -o tmp\_file} \\ 
        libjpeg-turbo-2.0.6 & \textit{djpeg @@} \\ 
        libpng-1.6.37 & \textit{libpng\_read\_fuzzer @@} \\
        bzip2-1.0.8 & \textit{bzip2 -dk @@} \\
        harfbuzz-2.8.0 & \textit{hb-draw-fuzzer @@} \\
        libxml2-2.9.10 & \textit{xmllint @@} \\
        tiff-4.2.0 & \textit{tiffcp -c lzma @@ tmp.tiff} \\
        \bottomrule
        \end{tabular}
        \subcaption{Programs used in our experiments}
        \label{tab:studied_programs}
    }
     \parbox{.5\textwidth}{
        \begin{tabular}{ll}
        \toprule
        \textbf{Fuzzer} & \textbf{Technical details} \\ 
        \midrule
        AFL \cite{afl} & Evolutionary search \\ 
        FairFuzz \cite{lemieux2018fairfuzz} & Evolutionary search + rare branches \\
        Neuzz \cite{she2019neuzz} & Evolutionary search + neural network \\
        MOPT \cite{lyu2019mopt} & Evolutionary search + particle swarm \\
        Angora \cite{chen2018angora} & Evolutionary search + taint tracking \\
        TortoiseFuzz \cite{wang2020not} & Evolutionary search + memory operation favor \\
        \bottomrule
        \end{tabular}
        \subcaption{State-of-the-art fuzzers used in our experiments}
        \label{tab:studied_fuzzers}
    }
    \caption{Experiment setup}
    \label{tab:experiment_setup}
\end{table}

\subsection{Edge coverage}

\begin{table}[h!]
    \centering
    \footnotesize
    \begin{tabular}{p{0.8cm}p{0.7cm}p{0.7cm}p{0.7cm}p{0.7cm}p{0.4cm}p{0.5cm}p{0.7cm}}
        \toprule
        \textbf{Program} & \textbf{Tortoise Fuzz} &\textbf{Angora} & \textbf{MOPT} & \textbf{FairFuzz} & \textbf{AFL} & \textbf{Neuzz} & \textbf{\tool}  \\
        \midrule
        \textit{djpeg} & 2383 & 2469 & \textbf{2602} & 2486 & 2386 & 2193 & 2569 \\
        \textit{xmllint} & 10916 & n/a & 11779 & 11718 & 11211 & 10601 & \textbf{12329} \\
        \textit{libpng} & 2743 & 2949 & 2958 & 2862 & 2774 & 2871 & \textbf{3041} \\
        \textit{bzip2} & 999 & n/a & 999 & 999 & 999 & n/a & \textbf{1278} \\
        \textit{tiffcp} & 6084 & 6834 & 7960 & 6578 & 6949 & 7729 & \textbf{8230} \\
        \textit{harfbuzz} & 9437 & \textbf{12178} & 10001 & 10025 & 9367 & 9829 & 9928 \\
        \textit{objdump} & 5177 & 6234 & 7377 & 5437 & 5326 & 5684 & \textbf{8363} \\
        \textit{size} & 2930 & \textbf{5528} & 4376 & 4245 & 2901 & 3075 & 5058 \\
        \textit{nm} & 3282 & 5566 & 5772 & 4817 & 3947 & 4630 & \textbf{6586} \\
        \textit{strip} & 4846 & \textbf{10086} & 8065 & 6473 & 5016 & n/a & 9457 \\
        \bottomrule
        \multicolumn{8}{l}{n/a indicates that we are unable to instrument programs or failed}\\
        \multicolumn{8}{l}{to run the fuzzer}\\
    \end{tabular}
    \caption{Detailed edge coverage of different fuzzers}
    \label{tab:edge_coverage}
\end{table}
To answer RQ1, we fuzz each program for 24 hours and 3 times. Then, we report the average results. Since each fuzzing process of Neuzz occupies an entire GPU (e.g., 10 GB) for 24 hours, we do not have sufficient GPU resources to run Neuzz multiple times. As a result, unlike other fuzzers, each program is fuzzed once with Neuzz. According to our experiment settings, Neuzz employs AFL for 1 hour to generate the initial training data, because of this, we only draw its last 23 hours. Note that this is also the same setting applied in their paper~\cite{she2019neuzz}. Although Angora and \tool~uses the same method to measure edge coverage, the number of covered edges counted in Angora and~\tool~are different given the same set of test inputs. To avoid this incomparable edge coverage problem, we run Angora's generated test inputs with \tool's instrumented binaries to re-compute edge coverage. By applying this conversion, only edge coverage achieved by Angora after 24 hours is computed and thus, a single value is plotted. Similar to Angora, we also convert TortoiseFuzz's edge coverage. The results are shown in Fig. \ref{fig:edge_coverage} and the accumulated edge coverage after 24 hours are listed in Table \ref{tab:edge_coverage}.

From Fig. \ref{fig:edge_coverage}, we observe that on 6 out of 10 programs, \tool~achieves the highest edge coverage. Especially, it outperforms other fuzzers with a large margin in programs {\it objectdump} and {\it nm}. For the remaining programs, \tool~is the second best fuzzer except {\it harfbuzz}. It is not surprise that Angora achieves relatively high edge coverage. Angora is benefited from a search algorithm based on gradient descent, which is similar to our hot-bytes mutation. Although Neuzz is guided by gradient-guided mutation, it is unable to defeat other AFL-liked fuzzers, i.e., MOPT and Fairfuzz. In two programs {\it djpeg} and {\it xmllint}, Neuzz is even the worst fuzzer. TortoiseFuzz achieves the lowest edge coverage in many programs (e.g., {\it libpng, tiffcp, nm, strip}) because its mutation strategy favors test inputs containing memory operations. Among 10 programs, {\it bzip2} has low stability (i.e., two runs with the same input may result in 2 different execution paths). Because of this, its edge coverage measurement could be inaccurate.

A closer investigation shows that, in most of the cases, \tool~typically starts with a low edge coverage, gradually goes up and then surpasses MOPT at around 15 hours. This is an expected behaviour as \tool~initially has limited guidance from the neural network and gradually builds on program-specific knowledge to evolve the seed pool towards covering uncovered branches. This optimization process is time-consuming. We do believe that \tool~will create even larger margin if the programs are fuzzed for a longer time. This is because we still see the upward trend of \tool~when the 24-hour limit is reached. 


According to our results, it is clear that the overall ranking of 7 fuzzers is \tool~$>$ Angora $>$ MOPT $>$ Fairfuzz $>$ Neuzz $>$ AFL $>$ TortoiseFuzz. Compared to Neuzz which simiarly uses neural networks to predict hotbytes, we achieve 1.2x more edge coverage across multiple target programs.

\begin{table}[h!]
    \centering
    \footnotesize
    \begin{tabular}{p{0.7cm}p{2.5cm}p{2.3cm}p{2.3cm}}
        \toprule
        \textbf{Program} & \textbf{Angora} & \textbf{Finch} & \textbf{MOpt} \\
        \midrule
        \textit{djpeg} & 2463 / 2478 / 6 &  2500 / 2607 / 49 & 2588 / 2621 / 14 \\
        \textit{harfbuzz} & 12008 / 12263 / 120 &  9897 / 9951 / 23 & 9856 / 10072 / 102     \\
        \textit{libpng} & 2947 / 2953 / 3 & 3013 / 3073 / 25 & 2955 / 2962 / 3         \\
        \textit{nm} & 5368 / 5790 / 173 & 6168 / 6837 / 298 & 5242 / 6094 / 377       \\
        \textit{objdump} & 5888 / 6662 / 321 & 8005 / 8655 / 270 & 7195 / 7543 / 142       \\
        \textit{size} & 5343 / 5629 / 131 & 4529 / 5341 / 375 & 4292 / 4535 / 113       \\
        \textit{strip} & 9309 / 10630 / 564 & 8865 / 10173 / 541 & 7857 / 8172 / 147       \\
        \textit{tiffcp} & 6791 / 6919 / 60 & 8097 / 8390 / 121 & 7923 / 8028 / 48        \\
        \bottomrule
    \end{tabular}
    \caption{Edge ranges of 3 top fuzzers across different runs}
    \label{tab:deviation}
\end{table}

To provide greater detail on the result of multiple runs, we show the ranges and standard deviations of 3 top fuzzers in the Table \ref{tab:deviation} where each cell has the format (min / max / std). It can be seen that the results across multiple runs are relatively stable.

\subsection{Branch distance bitmap}
\label{sect:branch_distance_bitmap}


\begin{table}[h!]
    \centering
    \footnotesize
    \begin{tabular}{lll}
        \toprule
        \textbf{Program} & \textbf{Edge coverage bitmap} & \textbf{Branch distance bitmap} \\ \midrule
        \textit{djpeg} & 2 & \textbf{7}\\
        \textit{xmllint} & \textbf{140} & 107\\
        \textit{libpng} & 13 & \textbf{56}\\
        \textit{tiffcp} & \textbf{353} & 227\\
        \textit{harfbuzz} & 33 & \textbf{46}\\
        \textit{objdump} & \textbf{298} & 237\\
        \textit{size} & 102 & \textbf{112}\\
        \textit{nm} & 69 & \textbf{104}\\
        \textit{strip} & 135 & \textbf{147}\\
        \bottomrule
    \end{tabular}
    \caption{Detailed edge coverage}
    \label{tab:bitmap_comparison}
\end{table}
\bigbreak
In Neuzz paper, a closely related work, its neural network is trained with edge coverage bitmap. Therefore, to answer RQ2, we compare edge coverage achieved from test inputs generated by two neural networks: (1) using branch distance bitmap and (2) using edge coverage bitmap. Both networks are trained with the same training data, which is the min-Pareto set of AFL's seed pool achieved after 24 running hours. Then, the entire seed pool is mutated using Algorithm \ref{algo:neu_mutate}. The details on the achieved edge coverage are shown in Table~\ref{tab:bitmap_comparison}.

From the table, we observe that branch distance bitmap has higher edge coverage than edge coverage bitmap in 6 out of 9 programs including \textit{djpeg}, \textit{libpng}, \textit{harfbuzz}, \textit{size}, \textit{nm} and \textit{strip}. We also notice that coverage bitmap is better than branch distance bitmap on programs containing many easy-to-cover branches which have not been covered. For example, it helps to cover 353 branches in  \textit{tiffcp} and 298 branches in \textit{objdump}, these number of covered branches are significantly larger than in other programs, e.g., 2 in \textit{djpeg} and 13 in \textit{libpng}. We believe that when not so many easy-to-cover branches remain (e.g., programs are executed for a longer time), our proposed branch distance bitmap will dominate its opponent, edge coverage bitmap, in term of guiding the mutation process to discover new branches. This is evidenced by the experiment results on RQ1.

It can be observed that the branch distance bitmap works better in most of the programs as expected. For some programs, however, if the edge coverage bitmap is able to offer useful guidance (e.g., when the covered and uncovered branches share the same hot-bytes), it may still be effective. The worst case scenario for edge coverage bitmap occurs when there are a small number of new covered branches. In other words, easy-to-cover branches are mostly covered and the machine learning model offers little information on what are the hot-bytes for covering the remaining branches. The results from Fig.~\ref{fig:edge_coverage} partially confirms our analysis. For most of the programs, while edge coverage achieved by Neuzz becomes fixed after 5 hours, \tool~maintains the upward trend throughout the whole process.

\subsection{Min-Pareto set}
\begin{figure*}[t!]
    \centering
    \includegraphics[width=0.80\textwidth]{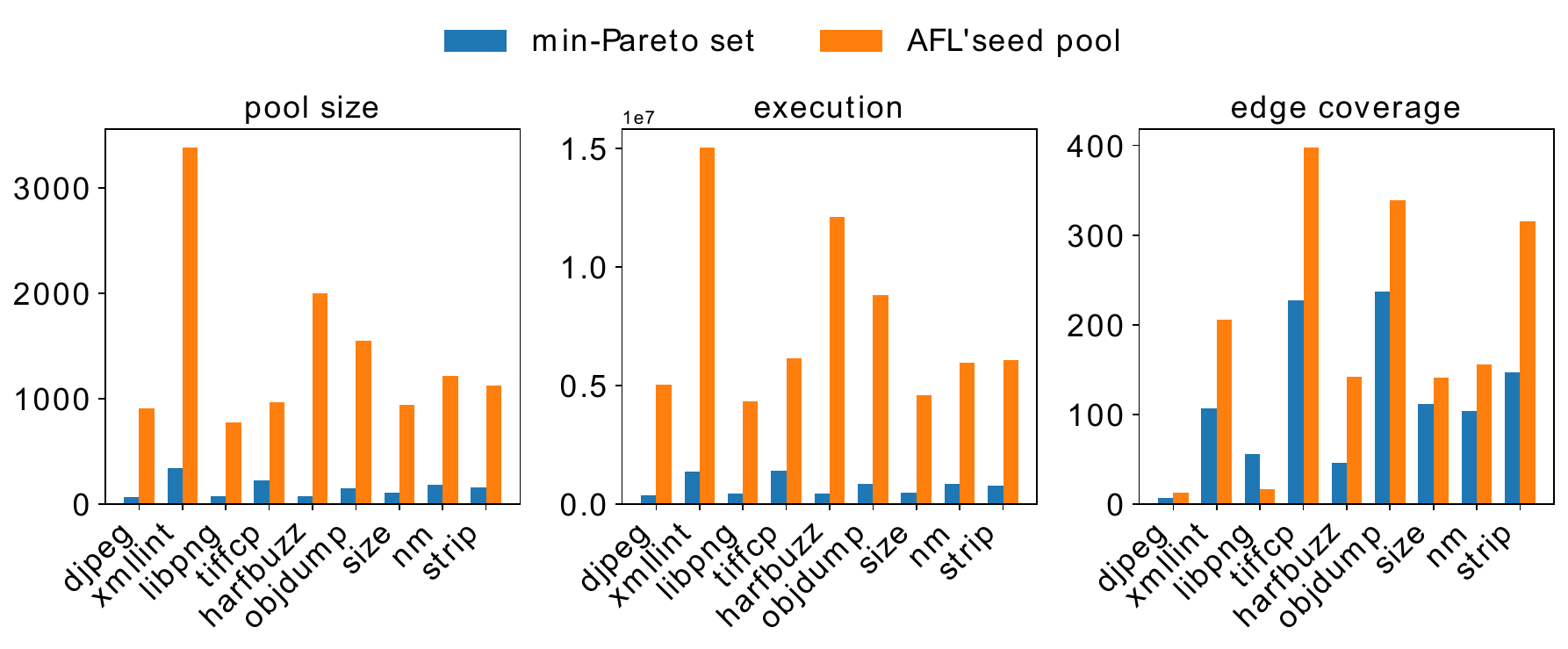}
    \caption{AFL's seed pool versus min-Pareto set}
    \label{fig:testsuite_comparison}
\end{figure*}
To answer RQ3, we reuse the training data of the experiments in Section \ref{sect:branch_distance_bitmap}. The neural network is trained with branch distance bitmap. We measure the number of test inputs of a seed pool a.k.a pool size, edge coverage and number of executions in two settings: (1) mutating the min-Pareto set and (2) mutating AFL's seed pool.

Fig. \ref{fig:testsuite_comparison} shows the summarized results from the two settings where x-axis represents target programs and y-axis is the unit of measured metrics. From the figure, we observe that our pool size is always much smaller than AFL's seed pool. On average, it accounts for 11.72\% of AFL's seed pool. Since a smaller pool results in a smaller number of generated test inputs and thus less executions, on average, our strategy executes 11.11\% test inputs but achieves 87.48\% edge coverage in comparison with AFL's seed pool. In some rare cases (e.g., {\it libpng}), with less test inputs, the min-Pareto set allows us discover more edges than mutating entire AFL's seed pool. This suggests that training neural network models with as many test inputs as possible is not always a good idea. This is because the training data is likely imbalanced, i.e., some paths are executed more often than others given the AFL's seed pool.

\subsection{Detecting bugs}

\definecolor{solarized@base03}{HTML}{002B36}
\definecolor{solarized@base02}{HTML}{073642}
\definecolor{solarized@base01}{HTML}{586e75}
\definecolor{solarized@base00}{HTML}{657b83}
\definecolor{solarized@base0}{HTML}{839496}
\definecolor{solarized@base1}{HTML}{93a1a1}
\definecolor{solarized@base2}{HTML}{EEE8D5}
\definecolor{solarized@base3}{HTML}{FDF6E3}
\definecolor{solarized@yellow}{HTML}{B58900}
\definecolor{solarized@orange}{HTML}{CB4B16}
\definecolor{solarized@red}{HTML}{DC322F}
\definecolor{solarized@magenta}{HTML}{D33682}
\definecolor{solarized@violet}{HTML}{6C71C4}
\definecolor{solarized@blue}{HTML}{268BD2}
\definecolor{solarized@cyan}{HTML}{2AA198}
\definecolor{solarized@green}{HTML}{859900}

\lstset{language=C++,
        basicstyle=\footnotesize\ttfamily,
        numbers=left,
        numberstyle=\footnotesize,
        tabsize=2,
        breaklines=true,
        escapeinside={@}{@},
        numberstyle=\tiny\color{solarized@base01},
        keywordstyle=\color{solarized@green},
        stringstyle=\color{solarized@cyan}\ttfamily,
        identifierstyle=\color{solarized@blue},
        commentstyle=\color{solarized@base01},
        emphstyle=\color{solarized@red},
        frame=single,
        rulecolor=\color{black},
        rulesepcolor=\color{black},
        showstringspaces=false,
        linewidth=0.45\textwidth
}
\begin{figure}[t]
    \centering
\begin{lstlisting}
unsigned int lava_get(unsigned int bug_num) {
  if (0x6c617661 - bug_num == lava_val[bug_num] ||
    SWAP_UINT32(0x6c617661 - bug_num) == lava_val[bug_num]) {
    printf("Successfully triggered bug %d, crashing now!\n", bug_num);
  }
  return lava_val[bug_num];
}
\end{lstlisting}
    \caption{A bug is triggered at function lava\_get()}
    \label{fig:function_lava_get}
\end{figure}

\emph{Artificial Dataset.} LAVA-M has a set of synthetic bugs inserted to 4 real-world programs {\it base64, md5sum, uniq, who} from project \textit{coreutils-8.24}. It is commonly used to measure the ability of discovering bugs of modern fuzzers \cite{chen2018angora, chen2019matryoshka, gan2020greyone}. In LAVA-M, a bug is identified by a unique number and guarded by a 4-byte comparison as shown in Figure \ref{fig:function_lava_get} (lines 2-3) where a bug is counted if the true-branch (lines 2-3) is satisfied. Considering a simple expression \textit{lava\_get(0)} + \textit{lava\_get(1)} + \textit{lava\_get(2)} leading to bugs 0, 1 and 2. Due to the deduplication mechanism of~\tool, whenever a bug is discovered (e.g., bug 0), remaining bugs (e.g., 1 and 2) will be discarded and not counted in our final results, even if they are triggered. This is because the remaining bugs do not achieve new coverage as bug 0 already bypassed the true-branch at line 2-3. To avoid such problem, we transform the above expression to \textit{hf0(0) + hf1(1) + hf2(2)} where functions {\it hf0, hf1, hf2} are identical to {\it lava\_get}. In the transformed version, the behavior of the original program is unchanged. However, each bug is associated with a unique branch and thus, the number of bugs is accurately counted in our final results. We apply this transformation to LAVA-M, conduct experiments and report the number of bugs found by~\tool~in Table \ref{tab:lava_m}. We compare our results with 7 other fuzzers where their number of bugs are taken from corresponding papers \cite{li2017steelix, chen2018angora, peng2018t, aschermann2019redqueen, gan2020greyone, chen2019matryoshka, she2019neuzz}. 

\tool~found the highest in number of bugs in 3 out of 4 programs. Especially, only \tool~is able to discover 61 bugs in program {\it md5sum}. Additionally, it triggers many unlisted bugs in LAVA-M. \tool~takes less than 1 hour to discover 48 and 29 bugs in \textit{md5sum} and \textit{uniq} respectively. After 6 hours, 61 bugs of \textit{md5sum} are exposed. For the slowest program \textit{who}, 2229 bugs are found within 24 hours. The outstanding results confirm that \tool~accurately selects the hot-bytes, mutate them to bypass magic-bytes comparisons.

\begin{table}[t]
    \centering
    \footnotesize
    \begin{tabular}{lllll}
        \toprule
         & base64 & md5sum & uniq & who \\
        \midrule
        \# Bugs & 44 & 57 & 28 & 2136 \\
        Fuzzer & 7 & 2 & 7 & 0 \\
        SES & 9 & 0 & 0 & 18 \\
        VUZZER & 17 & 1 & 27 & 50 \\
        Steelix* & 43 & 28 & 7 & 194 \\
        Angora & \textbf{48} & 57 & \textbf{29} & 1541 \\
        T-Fuzz & 43 & 49 & 26 & 63 \\
        RedQueen & 44 & 57 & 28 & 2134 \\
        GREYONE* & 44 & 57 & 28 & 2136 \\
        Matryoshka* & \textbf{48} & 57 & \textbf{29} & \textbf{2432} \\
        Neuzz & \textbf{48} & 60 & \textbf{29} & 1582 \\
        \tool & \textbf{48} & \textbf{61} & \textbf{29} & 2229 \\
        \bottomrule
        \multicolumn{4}{l}{* indicates that the fuzzer is not opensource}
    \end{tabular}
    \caption{Number of bugs found by different fuzzers on LAVA-M dataset}
    \label{tab:lava_m}
\end{table}

\noindent\emph{Real-world Applications.} Besides traditional crashes, to detect memory related bugs that does not necessarily cause a crash, we compile 10 real-world programs with ASAN. We run the same 6 fuzzers for 24 hours and report unique bugs found in Table \ref{tab:bug_found}. Note that the results are obtained after removing false positive and duplicated crashes. From Table \ref{tab:bug_found}, we observe that \tool~and Angora discover 1 previously unknown bug in program {\it tiffcp} and {\it strip} respectively, whereas all other fuzzers discover nothing.

\begin{table}[h!]
    \centering
    \footnotesize
    \begin{tabular}{p{0.8cm}p{0.7cm}p{0.7cm}p{0.7cm}p{0.8cm}p{0.3cm}p{0.7cm}p{0.7cm}}
        \toprule
        \textbf{Program} & \textbf{Tortoise Fuzz} & \textbf{Angora} & \textbf{MOPT} & \textbf{FairFuzz} & \textbf{AFL} & \textbf{Neuzz} & \textbf{\tool} \\
        \midrule
        \textit{tiffcp} & 0 & 0 & 0 & 0 & 0 & 0 & \textbf{1} \\
        \textit{strip} & 0 & \textbf{1} & 0 & 0 & 0 & 0 & 0 \\
        \textit{others} & 0 & 0 & 0 & 0 & 0 & 0  & 0\\
        \bottomrule
    \end{tabular}
    \caption{Unique bugs found by different fuzzers}
    \label{tab:bug_found}
\end{table}



\subsection{Training overhead}

\begin{figure}[t]
    \centering
    \includegraphics[width=0.45\textwidth]{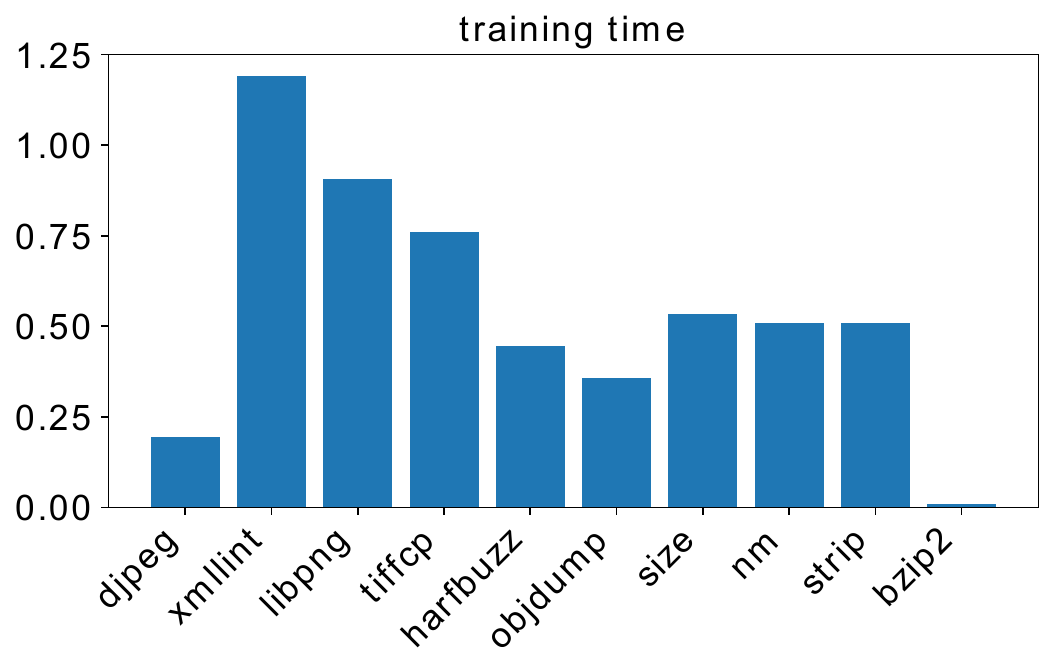}
    \caption{Training time of \tool~running for 24-hour experiments}
    \label{fig:time_overhead}
\end{figure}

To support hot-bytes mutation, \tool~introduces extra training time compared to AFL. Therefore, to answer RQ5, we measure the amount of time that the training task takes during our 24-hour experiments. Fig.~\ref{fig:time_overhead} summarizes the experiment results collected from 10 programs where y-axis is accumulated time in hours from multiple trainings during the lifetime of \tool. From Fig.~\ref{fig:time_overhead} we observe that in program {\it xmllint}, \tool~spends around 1.19 hours to train multiple min-Pareto sets. This is because {\it xmllint} is a complicated program which has many uncovered branches recorded as the training labels, e.g., the number of just-missed branches are 3449 when \tool~runs for 24 hours. In contrast, program {\it bzip2} is relatively small and thus the training time is insignificant. Intuitively, programs with many branches may result in more training time. On average, \tool~spends 0.54~hours for training. This amount of time is a reasonable trade-off to achieve better mutation results.
 
\section{Related Work}
\label{sect:related_work}
\tool~is closely related to machine learning based-fuzzers. Mohit Rajpal et al. \cite{rajpal2017not} employs multiple neural network architectures (e.g., LSTMs and sequence-to-sequence) to learn patterns in test inputs and guide the process of AFL's test input generation to reduce the number of useless mutations. Neuzz~\cite{she2019neuzz} uses gradient scores learned from a trained feed-forward neural network which approximates program's branching behaviours to assist the process of mutating hot-bytes (gradient-guided mutation). In \cite{she2019neuzz}, She et al. presented MTFuzz which is built on top of Neuzz. MTFuzz combines multiple training labels (i.e., edge-coverage, approach-sensitive and context-sensitive) to improve the accuracy of hot-bytes prediction. However, MTFuzz similarly suffers from the limitations of Neuzz. Machine learning techniques were also applied to different fuzzing steps, e.g., seed pool selection~\cite{wang2017skyfire,wang2019neufuzz} and test input mutation \cite{li2020v, bottinger2016deepfuzz}.

\tool~is complementary to existing fuzzers relying on search-based software testing approach. MooFuzz~\cite{math9030205} formulates test input selection and energy allocation as a many-objective optimization problem. Their mutated test inputs increase path risk and decrease energy consumption. sFuzz~\cite{nguyen2020sfuzz} is a smart contract fuzzer which employs a lightweight multi-objective adaptive strategy targeting those hard-to-cover branches to achieve new coverage. Essentially, it spend more effort on mutating a test input that is the best for a just-missed branch.

\tool~is related to fuzzers that analyze data flow dependency to identify promising mutating locations. Matryoshka~\cite{chen2019matryoshka} collects predicates in the conditional statement that guards uncovered branches and analyzes related control and data flow dependency to generate test inputs that can easily bypass constraints involving deeply nested conditional statements. Angora~\cite{chen2018angora} uses byte-level taint tracking to determine hot-bytes and bypass complicated constraints by mutating test inputs with the guidance of gradient descent algorithm. GREYONE \cite{gan2020greyone} infers taint of variables and uses this information to prioritize uncovered branches, bytes in the mutation algorithms. The data dependency analysis were also applied in many other fuzzers, e.g., REDQUEEN \cite{aschermann2019redqueen}, Vuzzer \cite{rawat2017vuzzer}, Steelix \cite{li2017steelix} and TaintScope \cite{wang2010taintscope}. Compared with the above-mentioned approaches, \tool~ quantifies the dependency with the help of a neural network trained with branch distances.

\tool~is broadly related to fuzzers that uses symbolic/concolic execution techniques to solve path constraints. SAVIOR \cite{chen2020savior} prioritizes the concolic execution of the seed that are likely to discover more vulnerabilities. Driller \cite{stephens2016driller} uses its fuzzing engine to explore most of the execution paths and switch to concolic execution only to bypass complicated conditions. T-fuzz \cite{peng2018t} fuzzes transformed programs to get out of stuck and leverages crash analyzers to remove false positives. Unfortunately, this approach often suffer from reliability and performance issues in practice.

\bigbreak
\vspace{-1cm}
\section{Conclusion}
\label{sect:conclusion}
In this work, we propose an approach to solve 2 fundamental problems in fuzzers, i.e., how to choose the seed and which bytes to mutate so that we can maximize the coverage as well as discover more bugs. Our approach is based on the idea that we can use the distances between test inputs and uncovered branches to choose the test inputs to mutate and to train a neuron network model which can help to guide the mutation process. The experiment results on popular programs show that our approach outperforms state-of-the-art tools. Currently, our implementation is for C/C++ programs. In future, we plan to extend our approach into a framework which can support the fuzzing of programs written in different languages.

\bibliographystyle{IEEEtran}
\bibliography{ref}
\end{document}